\documentclass[journal]{IEEEtranTIE}
\usepackage{graphicx}
\usepackage{cite}
\usepackage{picinpar}
\usepackage{amsmath}
\usepackage{url}
\usepackage{flushend}
\usepackage[latin1]{inputenc}
\usepackage{colortbl}
\usepackage{soul}
\usepackage{multirow}
\usepackage{pifont}
\usepackage{color}
\usepackage{alltt}
\usepackage[hidelinks]{hyperref}
\usepackage{enumerate}
\usepackage{siunitx}
\usepackage{breakurl}
\usepackage{epstopdf}
\usepackage{pbox}

\usepackage[tight,footnotesize]{subfigure}

\begin{document}
\title{Reliability of Power System Frequency \\ on Times-Stamping Digital Recordings}

\author{
	\vskip 1em
	
	Guang Hua, 
	Qingyi Wang,
	Dengpan Ye, 
	and Haijian Zhang, 

	\thanks{
	
		
		G. Hua, Q. Wang, and H. Zhang are with the School of Electronic Information, Wuhan University, Wuhan 430072, China (e-mail: ghua@whu.edu.cn; wqywqy@whu.edu.cn; haijian.zhang@whu.edu.cn). 
		
		D. Ye is with the School of Cyber Science and Engineering, Wuhan University, Wuhan 430072, China  (e-mail: yedp@whu.edu.cn).
	}
}

\maketitle
	
\begin{abstract}
Power system frequency could be captured by digital recordings and extracted to compare with a reference database for forensic time-stamp verification. It is known as the electric network frequency (ENF) criterion, enabled by the properties of random fluctuation and intra-grid consistency. In essence, this is a task of matching a short random sequence within a long reference, and the reliability of this criterion is mainly concerned with whether this match could be \emph{unique} and \emph{correct}. In this paper, we comprehensively analyze the factors affecting the reliability of ENF matching, including \emph{length of test recording}, \emph{length of reference}, \emph{temporal resolution}, and \emph{signal-to-noise ratio (SNR)}. For synthetic analysis, we incorporate the first-order autoregressive (AR) ENF model and propose an efficient time-frequency domain (TFD) noisy ENF synthesis method. Then, the reliability analysis schemes for both synthetic and real-world data are respectively proposed. Through a comprehensive study we reveal that while the SNR is an important external factor to determine whether time-stamp verification is viable, the length of test recording is the most important inherent factor, followed by the length of reference. However, the temporal resolution has little impact on the matching process. 
\end{abstract}


\definecolor{limegreen}{rgb}{0.2, 0.8, 0.2}
\definecolor{forestgreen}{rgb}{0.13, 0.55, 0.13}
\definecolor{greenhtml}{rgb}{0.0, 0.5, 0.0}

\section{Introduction}
\IEEEPARstart{P}{ower} system frequency, as an important parameter for grid monitoring, operation, and control, has also been widely exploited in security and forensic applications for its special properties \cite{Overview_Audio_Forensics_SPM}. First, due to the imbalance between generation and load, the frequency could not stay precisely at the nominal value of $50$ or $60$ Hz, and it instead exhibits as a random process fluctuating around the nominal value. Besides, the synchronization within an interconnected grid makes the fluctuations consistent among different nodes in the grid. These properties make the power system frequency a unique grid-specific and time-dependent signature. 

Meanwhile, the voltage signal from electric network could be captured unintentionally by digital audio and video \cite{Video_ENFVidDet} recordings if the device is directly connected to a power main or is in the proximity of electric activities. The instantaneous frequency of the captured electric network voltage signal is the so-called electric network frequency (ENF). To authenticate or verify the time-stamp of a suspect recording based on the ENF criterion, one could extract the captured ENF and match it to the corresponding frequency reference database, and the matched location indicates the true time-stamp. This implies that the availability of a long-range and high-quality frequency database is essential for such an application. 

\begin{figure}[!t]
	\centering
	\includegraphics[width=3.3in]{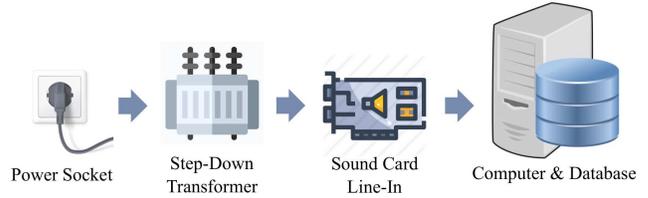}
	\caption{Demonstration of frequency data collection from power socket.}
	\label{database_demo}
\end{figure}

One representative example of power frequency reference database is the wide-area frequency monitoring network (FNET) \cite{FNET_Application} deployed north America, consisting of multiple frequency disturbance recorders (FDRs). This network collects high fidelity time-stamped frequency and other measurements, enabling tasks such as frequency prediction \cite{FNET_prediction}, grid data processing \cite{FNET_histogram}, monitoring and control \cite{FNET_distribution}, etc., besides serving as the forensic reference \cite{Experiment_PowerDel}.  Alternatively, according to intra-grid consistency, one may directly collect electric signal from an arbitrary power main using a step-down transformer and a PC, and the reference ENF could be obtained by estimating the signal?s instantaneous frequencies. A demonstrative example of such a lightweight reference ENF collection system is depicted in Fig. \ref{database_demo}, which could be deployed in any place with power supply.
Over the past decade, the ENF criterion has been extensively used in security-related applications including time-stamp verification \cite{Experiment_PowerDel,Own_DMA,App_BSim}, tampering detection \cite{Tamper_Phase,Tamper_IF,Tamper_Esprit_Hilbert,Own_AEM}, region of recording estimation \cite{App_Geo_Image,App_Region1}, and camera forensics \cite{App_Camera_Forensics}, while ENF extraction from suspect recordings has been the main technical challenge \cite{Estimation_ImproveDFT,Estimation_LP,Estimation_Harmonics2}. Recently, research focus has been put on practical issues, including factors affecting ENF capture in audio \cite{App_Factor_Audio} and video \cite{Video_Rolling_Shutter}. 

\subsection{Problem Statement and Motivation}
Recall the two important properties that enable the power system frequency to be a forensic criterion, i.e., intra-grid consistency and random fluctuation, the former has been well studied by Grigoras \cite{Experiment_RealCaseFull} in three different cities of Romania and then further addressed in subsequent works  from other parts of the world. It is also possible to design robust ENF reference databases \cite{App_Database}. For the latter, it has been discovered that the frequency fluctuation is approximately Gaussian distributed \cite{App_Gaussian_Model}, and Garg \emph{et al.} \cite{App_Gaussian_Model,Video_Seeing_ENF} have shown that the ENF could be modeled as a piecewise wide sense stationary (WSS) process.

Exploiting the ENF properties, a time-stamp verification process is considered successful if the extracted ENF from a suspect recording is \emph{\textbf{uniquely}} and \emph{\textbf{correctly}} matched to a piece of reference frequency segment, where the reference corresponds to a time interval (match scope) potentially containing the true time-stamp of test recording. By \emph{\textbf{uniquely}} it means that the matching process should yield a single best match via some measurements, e.g., mean squared error (MSE) or correlation coefficient (CC). By \emph{\textbf{correctly}} it means that the best match should indicate that the extracted test ENF and the reference segment are similar enough for us to make a decision, e.g., minimum MSE (MMSE) small enough to a certain satisfaction or CC close to one. 

\begin{table}[t!]
	\centering
	\caption{List of Commonly Used Recording Lengths}
	\vspace{-5pt}
	\renewcommand{\arraystretch}{1.3}
	\label{tab1}
	\begin{tabular}{c|c|c|r}
		\hline
		\hline
		Paper & Topic & Ref.? & \multicolumn{1}{c}{$L_\text{T}$}\\
		\hline
		\cite{Experiment_PowerDel} & Time-Stamp & Yes & $>10$ min\\
		\cite{Own_DMA} & Time-Stamp & Yes & $\mathbf{2\sim 15}$ \textbf{min}\\
		\cite{App_BSim} & Time-Stamp & Yes & $30$ min\\
		\cite{Tamper_Phase,Tamper_IF,Tamper_Esprit_Hilbert} & Tampering Detection & No & $\mathbf{19\sim 35}$ \textbf{s}\\	
		\cite{Own_AEM} & Tampering Detection & Yes & $6\sim 20$ min\\	
		\cite{Estimation_ImproveDFT} & Frequency Estimation & No & $\le  \mathbf{2}$ \textbf{min}\\
		\cite{Estimation_LP} & Frequency Estimation & Yes & $30$ min\\
		\cite{Estimation_Harmonics2} & Frequency Estimation & Yes& $50$ min\\
		\cite{App_Factor_Audio} & ENF Capture & No & $>10$ min\\
		\cite{Video_Seeing_ENF} & Video Forensics & Yes& $>10$ min\\
		\cite{Experiment_US} & Time-Stamp & Yes & $>40$ min\\
		\hline
		\hline
		\multicolumn{4}{l}{Note: ``Ref.?'' means whether reference ENF is needed.}
	\end{tabular}
\end{table}

Bearing this in mind, there naturally arises a question of how reliable is the time-stamp verification result given a test recording and a match scope. The confirmation of  frequency random fluctuation and intra-grid consistency is a prerequisite condition while the answer to this question is somewhat more complicated. It is implicitly agreed among the existing works that the length of recording should be more than $10$ min to ensure reliable matching \cite{Experiment_NetherlandsShort} (see summary in Table \ref{tab1}). However, in real-world situations a recording could have any length, and there lacks both theoretical and experimental support for $10$ min to be an appropriate threshold. This has given rise to a more specific question that till how short of a test recording will the ENF matching become unreliable. In fact, ENF reliability is determined not by a single factor, but by the combination of multiple factors including 

\begin{itemize}
	\item length of test recording ($L_\text{T}$);
	\item length of reference signal (match scope) ($L_\text{R}$);
	\item ENF estimation temporal resolution ($\delta$);
	\item signal-to-noise ratio (SNR).
\end{itemize} 

In this paper, we comprehensively investigate the above reliability-related factors. We concretize the concept of ENF reliability by answering the question of how would these factors affect the time-stamp matching result. We first study the AR modeling of the ENF, revealing its ultra low frequency (``frequency of frequency'') nature  and the ambiguity of ENF temporal resolution. Then, we propose an efficient AR($1$) model-based noisy ENF signal synthetic method directly in time-frequency domain (TFD), which could effectively avoid the unwanted influence of imperfect instantaneous frequency (IF) estimators. Real-world reference electric network signals of China Central Grid (recorded using the system depicted in Fig. \ref{database_demo}) are also used for the matching analysis. In addition, white Gaussian processes, whose fluctuation over time is strictly random, are considered for performance benchmark. Two analysis schemes are designed for real-world and synthetic test data respectively. Through a comprehensive and quantitative study, we discover the relationships among factors affecting ENF matching.  Note that in \cite{App_Factor_Video}, ENF-based time-stamp matching in video recordings has been analyzed. It focused specifically on light sources, video compression, and encoding methods. Our work differs from \cite{App_Factor_Video} by considering the above-mentioned inherent factors applicable to any form of recording media as long as the ENF is captured. In addition, our analysis is conducted in a more comprehensive and systematic manner.

The paper is organized as follows. ENF modeling and the  proposed noisy ENF synthesis method are detailed in Section II. Performance metrics and the analysis schemes for synthetic and real-world data are proposed in Section III respectively. In Section IV, quantitative analysis results of ENF reliability are presented, and the impacts of the considered factors are evaluated and summarized. Section V concludes the paper.

\section{ENF Modeling and Synthesis}
\subsection{AR Modeling of the ENF}
\subsubsection{The AR($1$) Model} All the involved signals in this paper are in discrete domain. The IF of electric network signal, i.e., the ENF, denoted by $f[n]$, is generated by passing a white Gaussian noise (WGN) denoted by $x[n]$, also termed as innovation signal, through an $M$-order infinite impulse response (IIR) filter. This is equivalent to an AR($M$) process. In the existing works, Garg \emph{et al.} \cite{App_Gaussian_Model} have set $M=1$ to study time-stamp verification, Hajj-Ahmad \emph{et al.} \cite{App_Region1} have set $M=2$ for ENF feature extraction, while Hajj-Ahmad \emph{et al.} \cite{Estimation_Compare} have set $M=1$ for a synthetic model. In this paper, we adopt the AR($1$) synthetic model, which is given by
\begin{equation}\label{ARM}
	f[n] =a f[n - 1] +x[n],
\end{equation}
where $a$ is the unknown coefficient, and $x[n]\sim\mathcal{N}(0,\sigma_x^2)$. Note that the mean value of the ENF is removed during ENF modeling \cite{App_Gaussian_Model}, thus $f[n]$ is a zero-mean process. For notation simplicity, we still use $f[n]$ to denote the ENF after the mean value is added back to it. The coefficient $a$ could be neatly solved via Levinson-Durbin recursion, with the use of the ground-truth frequency data. Our collection of the frequencyreference data from Central China Grid (for about $3$ years) is used for ENF modeling.

\begin{figure}[!t]
	\centering
	\includegraphics[width=3.4in]{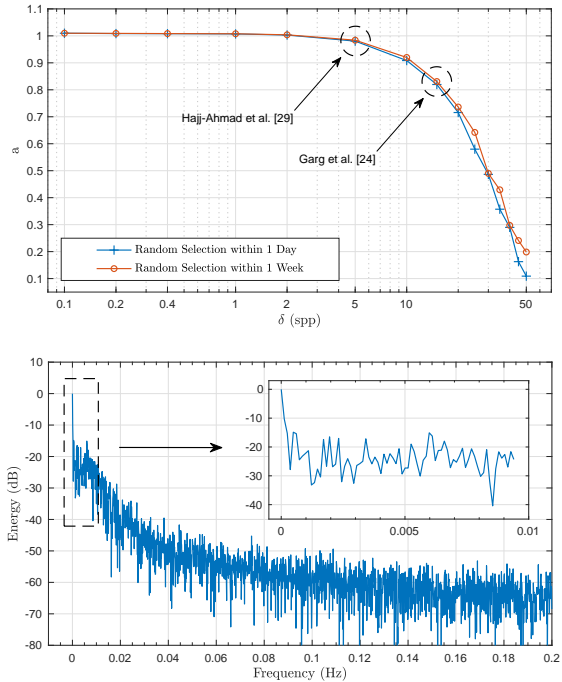}
	\caption{ENF modeling results. Upper: $a$ versus $\delta$. Lower: Energy density spectrum of a $2$-hour ENF reference, where $\delta=1$ spp.}
	\label{a_versus_delta}
\end{figure}

The model coefficient is determined together with the study of the temporal resolution of $f[n]$, denoted by $\delta$ second per point (spp). Consider the ENF $f[n]$ as a sampled discrete-time signal, then its sampling frequency is equal to $1/\delta$ Hz, which is the reciprocal of the temporal resolution. One should distinguish between the sampling frequency of time-amplitude domain (TAD) electric network signal and that of the TFD:
\begin{itemize}
	\item TAD sampling frequency: $f_\text{S}$ Hz;
	\item TFD sampling frequency: $1/\delta$ Hz, $1/\delta \le f_\text{S}$ .
\end{itemize}
To solve for $a$ and reveal the impact of $\delta$, we first randomly select a $1$-day (and $1$-month) TAD reference electric network signal. Then, a $1$-hour segment is randomly selected from the $1$-day (and $1$-month) reference, and the corresponding $f[n]$  is obtained with a resolution of $\delta$ spp. After that, $f[n]$ is fed into model (\ref{ARM}) to solve for $a$. The above process is repeated $100$ times, and the averaged values of $a$ as a function of $\delta$ are shown in Fig. \ref{a_versus_delta} upper plot. It can be seen when the temporal resolution is sufficiently high, i.e., $\delta<5$ spp, the AR($1$) model yields $a=1$. However, if $\delta>10$ spp, then it is unable to resolve the ENF and $a$ starts to drop rapidly. For the model to be stable (pole inside unit circle), we set $a=0.99$.

\subsubsection{Frequency of the ENF} The ENF is expected to be a very low frequency signal for its slowly-varying nature. To verify this, the energy density spectrum of a $2$-hour ENF reference segment is presented in Fig. \ref{a_versus_delta} lower plot, where $\delta=1$ spp. We observe that the energy concentrates at exteame low frequency region and becomes weaker than $-20$ dB when frequency is higher than $0.01$ Hz. Based on the two plots in Fig. \ref{a_versus_delta}, the frequency of the ENF in Central China Grid is approximately less than $1/20$ Hz, thus, the common setting $\delta=1$ spp is sufficient to sample the ENF (no aliasing). Note that Garg \emph{et al.} \cite{App_Gaussian_Model} have used $\delta=16$ spp, indicating the frequency of the ENF in the specific USA grid is less than $1/32$ Hz. 

\subsubsection{Temporal Resolution Ambiguity of AR(1)}
In the upper plot of Fig. \ref{a_versus_delta} it can be seen that if $\delta<2$ spp, then $a\to 1$. This indicates that we are unable to deduce $1/\delta$ or equivalently the TFD sampling frequency of the ENF given a synthetic $f[n]$ using  (\ref{ARM}) with $a=0.99$. We refer to this phenomenon as the temporal resolution ambiguity of AR($1$) model for the ENF. The reasons are twofold. First, AR($1$) may not be the optimal ENF model embodying all of its important statistical properties. This leads us to more in-depth study of ENF modeling, which is not in the scope of this paper. Second, high temporal resolution for ENF estimation may be a redundant setting, since the ENF is slowly varying and its frequency is very low as indicated by Fig. \ref{a_versus_delta}. 

\subsection{Synthesis of Noisy Electric Network Signal}
To reveal the inherent properties of the ENF, we exclude factors affecting ENF capture and base our analysis on the best situation in which the captured ENF is identical to the corresponding reference segment. Therefore, SNR-unrelated analysis is carried out by extracting a segment of the reference, considering it as the ENF extracted from a recording, and then matching it with a long reference signal. For SNR-related analysis, noise is added to the electric network signal before IF estimation. This requires us to synthesize noisy ENFs.

\begin{figure}[!t]
	\centering
	\includegraphics[width=3.3in]{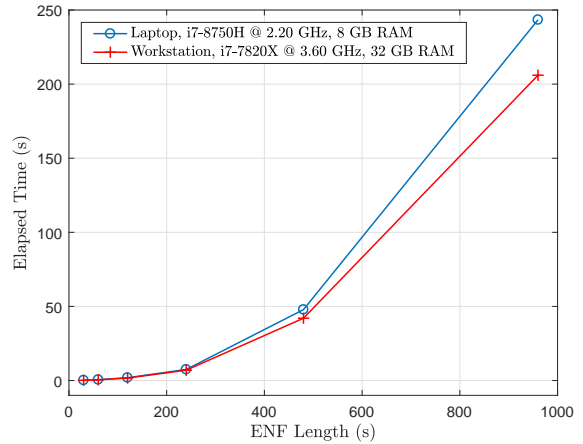}
	\caption{Elapsed time of TAD synthesis versus ENF length}
	\label{elapsed_time}
\end{figure}

\subsubsection{TAD Synthesis}
An intuitive way to synthesize a noisy ENF signal is to first synthesize the TAD waveform of the electric network signal, then add noise to the waveform, and finally perform IF estimation. The TAD electric network signal is modeled by a cosine wave with an unknown initial phase $\phi$  and time-varying amplitudes $A[n]>0$ and IFs, i.e., 
\begin{equation}\label{time_domain_model}
	s[n] = A[n]\cos \left( {2\pi T\sum\nolimits_{i = 0}^{n} {f[i]}  + \phi } \right),
\end{equation}
where $n=0,1,\ldots,N-1$,  the IFs $f[n]$ is the ENF, and  $T=1/f_\text{S}$ is the sampling interval. Denote the additive white Gaussian noise (AWGN) by $v[n]$, then we define the SNR by
\begin{equation}\label{SNR}
	\operatorname{SNR} = {{\sum\nolimits_n s^2[n]} \mathord{\left/
			{\vphantom {{\sum\nolimits_n s^2[n]} {\sum\nolimits_n v^2[n]}}} \right.
			\kern-\nulldelimiterspace} {\sum\nolimits_n v^2[n]}}.
\end{equation}
The synthetic noisy ENF is then obtained by the estimation of $f[n]$ from the noisy observation $s[n]+v[n]$ where the variance of $v[n]$ is determined by a given SNR. Based on the slowly varying nature of the ENF, the observed noisy signal is partitioned into overlapping frames and for each frame it is assumed the IF is constant within the frame interval. A single tone frequency estimator is then applied for each frame to extract the IF.  However, this process has  two limiations. 

\begin{itemize}
	\item The summation in (\ref{time_domain_model}) introduces a heavy computational burden, especially when $N$ is large. Specifically, to generate a single sample $s[n]$ at time $n$, the current and all the previous samples of $f[n]$ are summed. The total number of additions in (\ref{time_domain_model}) is $N(N+1)/2$. This is verified by Fig. \ref{elapsed_time}. The common synthesis of  $1$ week or longer noisy ENF thus becomes unrealistic even under current state-of-the-art computation power. 
	\item The imperfection of IF estimators prevents us from discovering the inherent properties of the ENF. It is known that Fourier analysis based normal resolution estimators are only asymptotically optimal, while high resolution estimators (e.g., MUSIC, ESPRIT) are very sensitive to noise. As a result, the inaccuracy of ENF matching may be caused not only by the said factors in Section I-B, but also by the gap between the selected estimator and a minimum variance unbiased estimator (MVUE).
\end{itemize}

\begin{table}[!t]
	\renewcommand{\arraystretch}{1.2}
	\addtolength{\tabcolsep}{-1pt}
	\caption{Scheme $1$ \--- Synthetic Data Analysis}
	\label{Scheme_Synthesis}
	\centering
	\vspace*{-6pt}
	\begin{tabular}{|rl|}
		\hline
		$1$. & Assign values to the factors:\\
		& $\cdot$ ENF temporal resolution $\delta$, $1/\delta \le f_\text{S}$,\\
		& $\cdot$ Search scope and reference length $L_\text{R}$,\\
		& $\cdot$ Test recording length $L_\text{T}$,\\
		& $\cdot$ SNR;\\
		$2$. & Synthesize $f_\text{R}[n]$ using (\ref{ARM}), $a=0.99$;\\
		$3$. & Obtain $f_\text{T}[n]$ as a random segment of  $f_\text{R}[n]$;\\
		$4$. & Record ground truth location $k_0$;\\
		$5$. & Synthesize noisy test ENF using (\ref{CRLB}) and (\ref{TFsynthesis});\\
		$6$. & Match $f_\text{T}[n]$ with $f_\text{R}[n]$ using (\ref{MMSE}) or (\ref{MCC});\\
		$7$. & Determine the tolerance parameter $\epsilon$;\\ 
		$8$. & Decide if the matching is successful using (\ref{decision}).\\
		\hline
	\end{tabular}
\end{table}

\subsubsection{TFD Synthesis}
To reduce the computational load and avoid the limitation of imperfect IF estimators, we propose a method to directly generate noisy ENF in the TFD. The variance of estimation errors caused by noise could be described by the Cramer-Rao lower bound (CRLB) given by \cite{StatisticalI}
\begin{equation}\label{CRLB}
	{{\mathop{\rm CRLB}\nolimits} _f} = \frac{{12}}{{{\mathop{\rm SNR}\nolimits}  \times {N_{\rm{F}}}\left( {N_\text{F}^2 - 1} \right)}} \times {\left( {\frac{{{f_{\rm{S}}}}}{{2\pi }}} \right)^2},
\end{equation}
where $N_\text{F}$ is the number of samples per frame. Assuming stationarity, the local SNR in (\ref{CRLB}) is approximated by the global SNR in (\ref{SNR}). Based on (\ref{CRLB}), the adding noise process, denoted by $g(f[n],\mathop{\rm SNR})$, is given by
\begin{equation}\label{TFsynthesis}
	g(f[n],{\mathop{\rm SNR}\nolimits} ) = f[n] + e[n],
\end{equation}
where $e[n]\sim\mathcal{N}(0,{{\mathop{\rm CRLB}\nolimits} _f})$. We could think of $g(f[n],\mathop{\rm SNR})$ as an MVUE of the ENF, because its mean and variance are
\begin{equation}
	\left\{\begin{array}{l}
		E\left\{ {g(f[n],{\mathop{\rm SNR}\nolimits} )} \right\} = f[n],\\
		{\mathop{\rm var}} \left\{ {g(f[n],{\mathop{\rm SNR}\nolimits} )} \right\} = {{\mathop{\rm CRLB}\nolimits} _f}.
	\end{array} \right.
\end{equation}
The proposal of (\ref{TFsynthesis}) for the synthesis of noisy ENF is much more efficient and accurate than first generating $s[n]$ followed by estimating the IFs from $(s[n]+v[n])$.

\section{Reliability Analysis Scheme}
\subsection{Performance Metrics}
We denote the test ENF by $f_\text{T}[n]$ and the reference ENF by $f_\text{R}[n]$, then the matched location, denoted by $\hat{k}$, is given by
\begin{equation}\label{MMSE}
	\hat k = \arg \mathop {\min }\limits_k \sum\nolimits_n {{{\left( {{f_{\text{T}}}[n] - {f_{\rm{R},k}}[n]} \right)}^2}},
\end{equation}
and 
\begin{equation}\label{MCC}
	\hat k = \arg \mathop {\max }\limits_k \frac{{\sum\nolimits_n {\left[ {\left( {{f_{\text{T}}}[n] - {\mu _{\text{T}}}} \right)\left( {{f_{{\text{R}},k}}[n] - {\mu _{{\text{R}},k}}} \right)} \right]} }}{{\sqrt {\sum\nolimits_n {{{\left( {{f_{\text{T}}}[n] - {\mu _{\text{T}}}} \right)}^2}\sum\nolimits_n {{{\left( {{f_{{\text{R}},k}}[n] - {\mu _{{\text{R}},k}}} \right)}^2}} } } }},
\end{equation}
based on the MSE and CC respectively, where ${f_{\text{R},k}}[n]=f_\text{R}[n+k]$, $\mu_\text{T}$ is the sample mean of $f_\text{T}[n]$, and $\mu_{\text{R},k}$ is the sample mean of ${f_{\text{R},k}}[n]$. In most situations the two metrics yield very similar results. The former is more computationally efficient while the latter is slightly more accurate. In our study, the two metrics are used alternatively where appropriate.  Further, let the ground truth matched location be $k_0$, then the absolute matching error is given by $|\hat{k}-k_0|$. We consider the matching result to be successful if 
\begin{equation}\label{decision}
	\big|\hat{k}-k_0\big|< \epsilon,
\end{equation}
where $\epsilon$ is a toleration parameter. This means that matching errors within $\epsilon\delta$ seconds are acceptable in practice. 

\begin{table}[!t]
	\renewcommand{\arraystretch}{1.2}
	\caption{Scheme $2$ \--- Real-World Data Analysis}
	\label{Scheme_Real}
	\centering
	\vspace*{-6pt}
	\begin{tabular}{|rl|}
		\hline
		$1$. & Assign values to the factors:\\
		& $\cdot$ ENF temporal resolution $\delta$, $1/\delta \le f_\text{S}$,\\
		& $\cdot$ Search scope and reference length $L_\text{R}$,\\
		& $\cdot$ Test recording length $L_\text{T}$,\\
		& $\cdot$ SNR;\\
		$2$. & Obtain TAD reference ENF from database;\\
		$3$. & Select TAD test ENF randomly from reference;\\
		$4$. & Record and convert ground truth location to $k_0$;\\
		$5$. & Add noise to the test signal according to SNR;\\	
		$6$. & Estimate $f_\text{T}[n]$ and $f_\text{R}[n]$ based on STFT;\\
		$7$. & Match $f_\text{T}[n]$ with $f_\text{R}[n]$ using (\ref{MMSE}) or (\ref{MCC});\\
		$8$. & Determine the tolerance parameter $\epsilon$;\\ 
		$9$. & Decide if the matching is successful using (\ref{decision}).\\
		\hline
	\end{tabular}
\end{table}

\subsection{Proposed Analysis Schemes}
The analysis schemes are summarized in Tables \ref{Scheme_Synthesis} and \ref{Scheme_Real} for synthetic data (Scheme $1$) and real-world data (Scheme $2$) respectively. The schemes are parameterized by the factors under investigation. Note that Scheme $1$ directly operates in TFD, thus time-frequency analysis is not needed. However, time-frequency analysis is need in Scheme $2$ to transform the real-world data from TAD to TFD. We adopt the widely used frequency estimator which combines the short-time Fourier transform (STFT) and an interpolation mechanism \cite{Estimation_Compare}, for its balance among time/frequency resolution, robustness against AWGN, and computational complexity.

In both schemes, the same factors are initialized in Step $1$. In Step $2$ of Scheme $1$, $\delta$ is the sampling interval of the innovation signal $x[n]$, while in Step $6$ of Scheme $2$, $\delta$ is the STFT frame step-size. Besides, the test ENF is in fact a random segment of the reference ENF, where the reference is either synthesized in TFD from the AR($1$) model or obtained in TAD from the reference database. In this way, one can exclude unwanted factors, e.g., potential problems caused by test recording devices, non-stationary noise, clock desynchronization, etc., and focus on the inherent factors affecting ENF matching.

In addition, to gain more insights about the reliability of the frequency fluctuation as a forensic criterion, we also use independent and identically distributed (i.i.d.) Gaussian signal as the ENF. It represents the case of the most random fluctuation. Since such a signal is strictly random in which even consecutive samples are uncorrelated, the temporal resolution $\delta$ becomes a trivial factor. This analysis is realized by replacing Step $2$ of Scheme $1$ with a white Gaussian distributed sequence of a length defined by the match scope.

\begin{figure}[!t]
	\centering
	\includegraphics[width=3.3in]{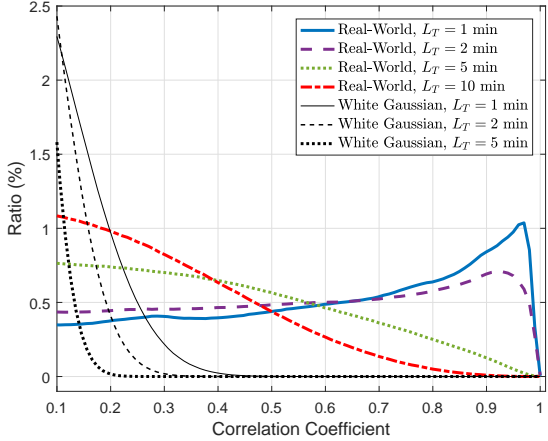}
	\caption{Ratio of different CC values for ENF matching in noise-free situation, averaged from $100$ random realizations, where $f_\text{S}=400$ Hz, $\delta = 1$ spp, and length of reference $L_\text{R}=168$ h ($1$ week).}
	\label{CC_Count}
\end{figure}
\section{Results and Discussions}
This section presents the analysis results on the factors affecting forensic ENF matching, i.e., $L_\text{T}$, $L_\text{R}$, $\delta$, and SNR, respectively. Experiments using synthetic data are carried out via Scheme $1$ while those using real-world reference ENF data are carried out via Scheme $2$. 

\begin{figure}[!t]
	\centering
	\includegraphics[width=3.3in]{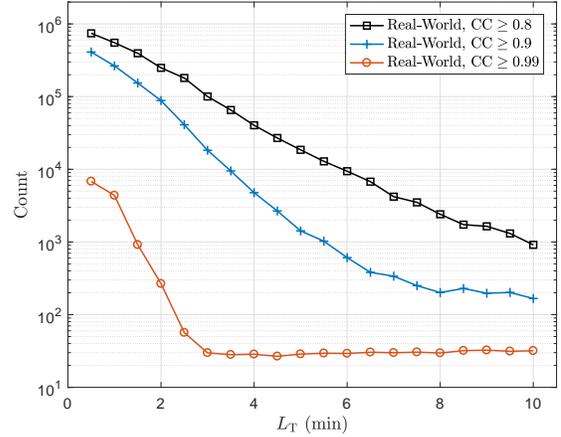}
	\caption{Counts of similar matches versus $L_\text{T}$ in noise-free situation, averaged from $100$ random realizations where $f_\text{S}=400$ Hz, $\delta = 1$ spp, and $L_\text{R}=168$ h ($1$ week).}
	\label{CC_Count_Ratio}
\end{figure}

\begin{figure}[!t]
	\centering
	\includegraphics[width=3.3in]{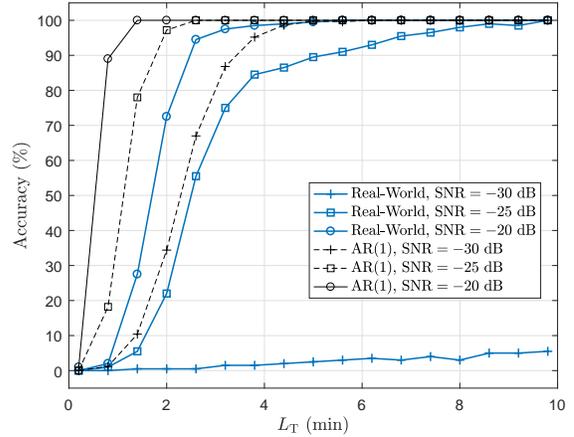}
	\caption{Analysis results on the factor of length of test ENF, i.e., $L_\text{T}$, averaged from $100$ random realizations, where $f_\text{S}=400$ Hz, $\epsilon=10$ s, $\delta=1$ spp, and $L_\text{R} = 168$ h ($1$ week).}
	\label{Acc_LT}
\end{figure}

\subsection{Length of Test Recording}
We first consider the ideal noise-free case, in which we choose an arbitrary $168$-hour ($1$ week) reference ENF and extract a small segment as the test data. Due to the random fluctuation property, there would be exactly one matched location corresponding to zero MSE. However, there would exist similar matches in other locations. This is illustrated in Fig. \ref{CC_Count}, where CC is used as the metric and the region $\text{CC} \in [-1, -0.1)$ is not presented for clarity. It can be seen that for short recordings, e.g., $L_\text{T}=1$ or $2$ min, there exist a large ratio of reference segments having CC values close to $1$. This indicates that in practical situations where noise is inevitable, those similar segments would introduce false matches. As the length of test recording increases, the amount of similar matches is reduced rapidly. Noticeably, for white and Gausian distributed benchmark signal, it exhibits high uniqueness property as there is almost no match yielding a CC value greater than $0.5$ even when $L_\text{T}=1$ min. The averaged counts of similar ENF matches quantified by $\text{CC}\ge 0.8$, $0.9$, and $0.99$ respectively are presented in Fig. \ref{CC_Count_Ratio} for different values of $L_\text{T}$. It shows that for a match scope spanning $1$ week, even for long test recordings of $10$ min, there exist about $1000$ reference segments yielding $\text{CC}\ge 0.8$, $200$ yielding $\text{CC}\ge 0.9$, and $30$ yielding $\text{CC}\ge 0.99$. As $L_\text{T}$ reduces, the numbers of similar matches increase significantly. When $L_\text{T}<3$ min, even the number of similar matches with $\text{CC}\ge 0.99$ increases rapidly.

We now consider noisy situations with AWGN, and results are shown in Fig. \ref{Acc_LT}. Due to the weakness of the electric network signal compared to recorded audio content and noise, we consider relatively low SNR situations here. We note from this figure that $\text{SNR}=-30$ dB leads to the failure of ENF matching, while the synthetic signal does not suffer from this. Further, the accuracies increase rapidly when $L_\text{T}>2$ min, indicating that when the match scope is $1$ week, the length of test recording is preferred to be greater than $2$ min for reliable forensic matching. However, in high SNR situations, one may still successfully work with reduced test recording lengths. 

\begin{figure}[!t]
	\centering
	\includegraphics[width=3.2in]{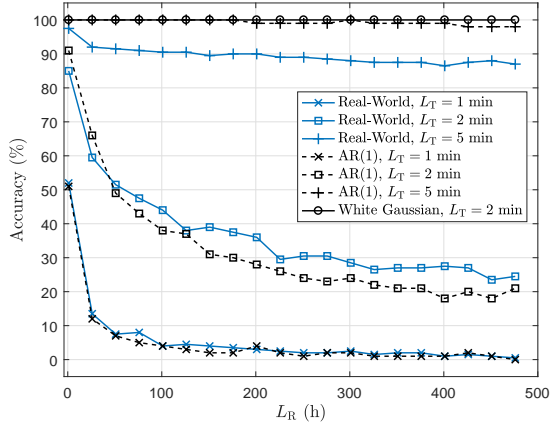}
	\caption{Analysis results on the factor of length of reference ENF, i.e., $L_\text{R}$, averaged from $100$ random realizations, where $f_\text{S}=400$ Hz, $\epsilon=10$ s, $\delta=1$ spp, and $\text{SNR}=-25$ dB.}
	\label{Acc_LR}
\end{figure}

\subsection{Length of Reference ENF (Match Scope)}
We consider a wide range of match scopes from $1$ to $496$ h ($\approx 3$ weeks). The matching accuracies versus the match scope is presented in Fig. \ref{Acc_LR} using real-world reference, AR($1$) synthetic data, and white Gaussian signals respectively. When $L_\text{T}= 1$ and $2$ min, the accuracy curves of the real-world and AR($1$) synthetic data are very close to each other. They both decrease rapidly when $L_\text{R}$ increases from $1$ to $50$ h. This indicates that when dealing with very short recordings, it is necessary to narrow down the match scope to be less than $2$ days to ensure reliable matching. For $1$-min test recordings and under $-25$ dB SNR, the accuracy only reaches $50\%$ even when $L_\text{R}=1$ h. The only way to improve matching accuracy for this challenging case may be trying to suppress the noise and improve the SNR condition, as indicated in Fig. \ref{Acc_LT}. We further note from Fig. \ref{Acc_LR} that the accuracies are substantially improved for longer test recordings. When $L_\text{T}=5$ min, the accuracies of real-world and AR($1$) data reach the levels of $90\%$ and $100\%$ respectively. For white Gaussian signal, the length of reference data is not an issue as we can see that the accuracies are constantly $100\%$ across the values of $L_\text{R}$ even when the duration of test ENF is only $2$ min.

\begin{figure}[!t]
	\centering
	\includegraphics[width=3.2in]{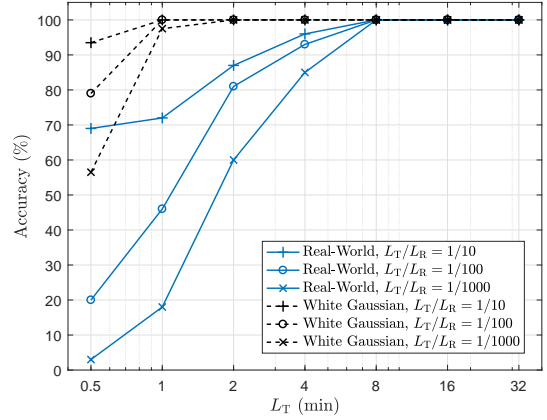}
	\caption{Analysis results on the influence of $L_\text{T}/L_\text{R}$ on matching accuracy, averaged from $100$ random realizations, where $f_\text{S}=400$ Hz, $\epsilon=10$ s, $\delta=1$ spp, and $\text{SNR}=-25$ dB.}
	\label{Acc_LT_LR_Ratio}
\end{figure}

To provide more insights, we further look into the effects of the ratio $L_\text{T}/L_\text{R}$, and the results are presented in Fig. \ref{Acc_LT_LR_Ratio}. Intuitively, for short test recordings, one may work to narrow down the match scope so that the test ENF is more likely to be correctly matched to the reference data; for long test recordings, the match scope could be reasonably longer. However, it can be seen from Fig. \ref{Acc_LT_LR_Ratio} that the accuracies are not a constant for a fixed ratio of $L_\text{T}/L_\text{R}$. It is observed that for a $0.5$-min real-world test ENF, the accuracy is only $70\%$ even if  $L_\text{R}=5$ min. If $L_\text{R}=5$ is increased to $50$ min, i.e., $L_\text{T}/L_\text{R}=1/100$, the accuracy becomes only $20\%$. On the other side, if $L_\text{T}\ge 8$ min, the matching accuracies become $100\%$ for both real-world and white Gaussian data even if $L_\text{T}/L_\text{R}=1/1000$. For a $32$-min test ENF, the accuracy is $100\%$ even if the match scope spans $23$ days.

\begin{figure}[!t]
	\centering
	\includegraphics[width=3.2in]{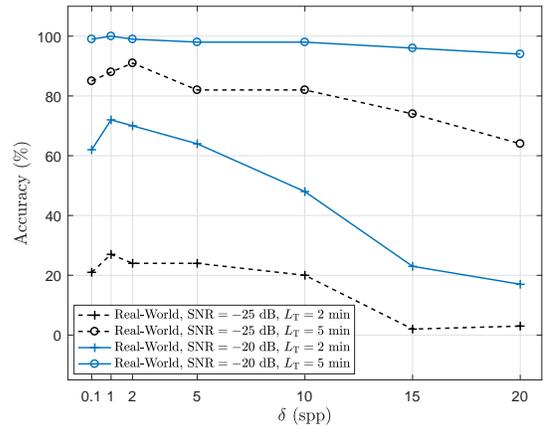}
	\caption{Analysis results on the factor of ENF temporal resolution $\delta$ using real-world data, averaged from $100$ random realizations, where $f_\text{S}=400$ Hz, $\epsilon=10$ s, and $L_\text{R} = 168$ h ($1$ week).}
	\label{delta}
\end{figure}

\subsection{Temporal Resolution $\delta$}
We now investigate the influence of ENF temporal resolution $\delta$ on the matching accuracy. Due to the temporal resolution ambiguity of synthetic data, results shown in this subsection are based on real-world data only. The matching accuracies under different values of $\delta$ are presented in Fig. \ref{delta}. It can be seen that the accuracies are generally not very sensitive to $\delta$ as we observe that the curves do not intersect. Noticeably, when $\text{SNR}=-20$ dB and $L_\text{T}=5$ min, the degradation is minimal ($<5\%$) even if $\delta = 20$ spp. This result is generally consistent with the analysis of the frequency of the ENF presented in Section II-A-2). Meanwhile, it is important to note that using very high temporal resolution (e.g., $\delta=0.1$ spp) may also cause performance degradation. Another drawback of using very high temporal resolution is the linearly increased computational burden. Therefore, a sound choice of $\delta$ should ensure no loss of ENF detail while maintaining a reasonable computational complexity. According to Fig. \ref{delta}, a sound choice of ENF temporal resolution could be $\delta=1$ to $5$ spp.

\begin{figure}[!t]
	\centering
	\subfigure[Real-world data.]{\includegraphics[width=3.2in]{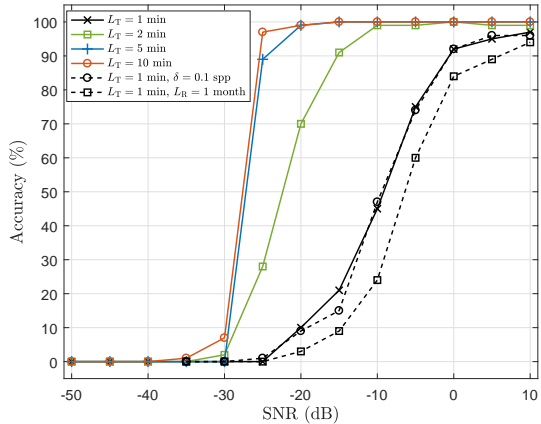}}\\[-3pt]
	\subfigure[Synthetic data.]{\includegraphics[width=3.2in]{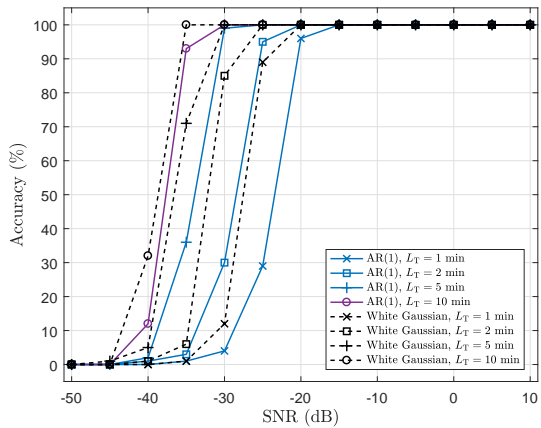}}
	\caption{Analysis results on the factor of SNR using real-world and synthetic data, averaged from $100$ random realizations, where $f_\text{S}=400$ Hz, $\epsilon=10$ s, $\delta=1$ spp, and $L_\text{R} = 168$ h ($1$ week).}
	\label{Acc_SNR}
\end{figure}

\subsection{SNR}
Matching accuracies under different SNR values are shown in Figs. \ref{Acc_SNR}. We observe that for real-world data, $\text{SNR}=-30$ dB is again seen as the boundary under which none of the settings could work. However, this boundary reduces to $-40$ dB for the AR($1$) and white Gaussian models. In Fig. \ref{Acc_SNR} (a), the test ENF exhibits consistently high uniqueness when $L_\text{T}>5$ min. $L_\text{T}$ is thus again seen to be the most important factor in ENF matching. Besides, for the $1$-min test recording, increasing the temporal resolution by $10$ times did not offer any performance gain, while increasing $L_\text{R}$ from $1$ week to $1$ month resulted in noticeable degradation. However, reliable ENF matching could still be expected if $\text{SNR}>10$ dB. In Fig. \ref{Acc_SNR} (b), both the curves using AR($1$) and white Gaussian signals could serve as the performance bounds, but they are somewhat loose as compared with the curves in Fig. \ref{Acc_SNR} (a). 

\subsection{Summary and Discussion}
In summary, the four factors determining ENF-reliability are ranked according to their impacts in descending order, as shown in \ref{tab4}. Among these factors, SNR is an external one, while the others reflect inherent ENF properties. If the SNR prevents us from extracting the ENF from the recording (e.g., $\text{SNR}<-30$ dB), then ENF analysis becomes inapplicable. 

If the SNR permits ENF-based forensic analysis, then among the other factors, the length of test ENF] $L_\text{T}$ has the most significant impact. Based on results shown in Figs. \ref{Acc_LT}\--\ref{Acc_LT_LR_Ratio} and \ref{Acc_SNR} (a), if $L_\text{T}<5$ min, then it simultaneously requires a high SNR condition and a narrow match scope for us to have a successful match. It is important to specially note for very short recordings with $L_\text{T}<1$ min, which requires a very high SNR and an extremely narrow match scope. However, if the test ENF is sufficiently long, e.g., $L_\text{T}>10$ min, then it is highly possible to correctly match the time-stamp even if $L_\text{R}$ is longer than $1000$ times of $L_\text{T}$, as seen from Figs. \ref{Acc_LR}\--\ref{Acc_SNR}.

While $L_\text{T}$ is solely determined by who made the recording, it is possible to work by various means to narrow down the match scope $L_\text{R}$. If $L_\text{R}$. contains the true time-stamp information, then a smaller $L_\text{R}$ generally yields improved matching accuracy. However, the matching accuracy is not a constant for a fixed test-reference ratio $L_\text{T}/L_\text{R}$, as shown in Fig. \ref{Acc_LT_LR_Ratio}. Therefore, if $L_\text{T}$ is the primary factor that determines how successful it could be matched to the reference data, then $L_\text{R}$ serves as a secondary one, reducing it could further improve matching accuracy.

\begin{table}[t!]
	\centering
	\caption{Summary of Factors Affecting ENF Reliability (Ranked)}
	\vspace{-5pt}
	\renewcommand{\arraystretch}{1.20}
	\label{tab4}
	\begin{tabular}{c|c|c|c}
		\hline
		\hline
		Rank & Factor & Nature & Preferable Condition\\
		\hline
		$1$ & SNR & External & $\text{SNR} > -25$ dB\\
		$2$ & $L_\text{T}$ & Inherent & $L_\text{T}>5$ min, longer the better\\ 
		$3$ & $L_\text{R}$ & Inherent & $L_\text{R}>L_\text{T}$, shorter the better\\
		$4$ & $\delta$ & Inherent & $1\le \delta \le 5$ spp\\
		\hline
		& \multicolumn{3}{c}{Impact Relationship: $\text{SNR}>L_\text{T}\gg L_\text{R}\gg\delta$}\\
		\hline
		\hline
	\end{tabular}
\end{table}

For the temporal resolution $\delta$, we have found that it has little impact on the matching accuracy. On the one hand, the slowly varying nature of the ENF yields its frequency to be lower than $1/20$ Hz in Central China Grid, as discussed in Section II-A-2). This indicates that as long as the TFD ENF sampling frequency $1/\delta>1/10$ Hz, i.e., $\delta<10$ spp, there is no loss of ENF information according to Nyquist theorem. On the other hand, a too granular ENF signal, e.g., $\delta=0.1$ spp, may cause negative effect, as shown in Fig. \ref{delta}. Ultra high temporal resolution may be redundant in forensic ENF matching, which is somewhat opposite to our intuition.

\section{Conclusion}
Power system frequency could be utilized to forensically time-stamping audio recordings, but there is a lack of research to answer under what conditions could the time-stamp matching be uniquely and accurately performed. This paper has answered this question in a comprehensive manner, considering a set of four factors. We have proposed a TFD noisy ENF synthesis method which is more accurate
and computationally efficient than the direct TAD synthesis. Using the proposed analysis schemes for both real-world and synthetic data, we have come to the conclusion that the SNR is the most important external factor. Among the other three inherent factors, the length of test recording is the most important factor, followed by the length of reference ENF (search scope). In addition, the ENF matching process has been found to be insensitive to ENF temporal resolution. This paper reveals the challenge in dealing with very short recordings (e.g., $L_\text{T}<1$ min).  Besides, the AR($1$) model may not fully characterize the statistical properties of the ENF. Note that AR($1$) with $a\to 1$ is approximately a random walk model, which is non-stationary. However, the ENF is in fact seen to be a stationary process. Therefore, it is worthy to carry out in-depth research on ENF modeling for future work.


\bibliographystyle{IEEEtranTIE}
\bibliography{ENFrefs}

\end{document}